# FloodGenome: Interpretable Machine Learning for Decoding Features Shaping Property Flood Risk Predisposition in Cities

Chenyue Liu[1*], Ali Mostafavi[2]

[1]Ph.D. Student, Urban Resilience.AI Lab, Zachry Department of Civil and Environmental Engineering, Texas A&M University, College Station, United States; e-mail: liuchenyue@tamu.edu

[2]Associate Professor, Urban Resilience.AI Lab Zachry Department of Civil and Environmental Engineering, Texas A&M University, College Station, United States; e-mail: amostafavi@civil.tamu.edu

## Abstract

Understanding the fundamental characteristics that shape the inherent flood risk disposition of urban areas is critical for integrated urban design strategies for flood risk reduction. Flood risk disposition specifies an inherent and event-independent magnitude of property flood risk and measures the extent to which urban areas are susceptible to property damage if exposed to a weather hazard. This study presents FloodGenome as an interpretable machine learning model for evaluation of the extent to which various hydrological, topographic, and built-environment features and their interactions shape flood risk disposition in urban areas. Using flood damage claims data from the U.S. National Flood Insurance Program covering the period 2003 through 2023 across four metropolitan statistical areas (MSAs), the analysis computes building damage ratios and flood claim counts by employing k-means clustering for classifying census block groups (CBGs) into distinct property flood risk disposition levels. Then a random forest model is created to specify property flood risk levels of CBGs based on various intertwined hydrological, topographic, and built-environment features. The transferability of the model in assessing the determinants of flood risk disposition across different MSAs is evaluated based on training a random forest model on data from one MSA and then applying it to other MSAs. The model transferability analysis results show consistent performance across MSAs, revealing the universality of underlying features that shape city property flood risks. Feature importance analysis results reveal that average annual vapor pressure, height above nearest drainage (HAND), elevation above sea level, and imperviousness are the primary features shaping the property flood risk disposition of urban areas. Given the universality of features shaping flood risk disposition of urban areas, the FloodGenome model is then used to: (1) evaluate the extent to which future urban development would exacerbate flood risk disposition of urban areas; and (2) specify property flood risk levels at finer spatial resolution providing critical insights for flood risk management processes. The FloodGenome model and the findings provide novel tools and insights for improving the characterization and understanding of intertwined features that shape flood risk profiles of cities.

These outcomes are essential for flood managers and urban planners to develop integrated urban design strategies that alleviate flood risk disposition of urban areas.

# 1 Introduction

The objective of this study is to unveil intertwined features that shape property flood risk predisposition in cities. Property flood risk disposition specifies an inherent and event-independent magnitude of property flood risk which measures the extent to which urban areas are susceptible to property damage if exposed to weather hazards. Such event-independent evaluation of flood risk disposition is analogous to evaluating the genetic factors that shape individuals' predisposition to inherited diseases. An area with a greater flood risk disposition would experience greater property damage when a weather hazard event occurs in the area. The features that shape flood risk predisposition evolve due to urban development and growth. Hence, the evaluation of the underlying features that shape property flood risk disposition is essential for informing urban development plans and integrated urban design strategies that alleviate flood risk propensity of an area.

This majority of urban flood risk assessment efforts are fueled by the motivation for estimating inundation extent [1] [2] [3] and its associated likelihood at the finest possible resolution [4]. This pursuit is primarily propelled by risk management strategies such as flood insurance programs and their desire to quantify flood exposure. However, from an urban planning and urban flood risk management perspective, it is essential to evaluate the extent to which various urban features (influenced by development patterns and growth) would change the inherent risk of spatial areas to property flood risk. While the existing physics-based hydraulic and hydrologic models [5] capture the underlying features that influence flood inundation, due to their limited interpretability, the outcomes provide inundation depth levels with little to no explainability of the relative importance of the underlying features and their interactions in shaping flood extent. Also, incorporating changes in urban development patterns into physics-based models is rather effort- and computing-intensive, limiting its utility for evaluating ways changes in urban development influence property flood risk disposition of urban areas. Recognizing this gap, in this study, we create an explainable machine learning model to evaluate how various hydrological, topographic, and built-environment features shape property flood risk disposition.

The application of machine learning-based methods has grown in recent years. Machine learning (ML) algorithms, such as random forest [6] [7], decision trees [8] and support vector machines [9], are increasingly applied to flood risk prediction. The emergence of deep learning (DL) enhances flood predictive monitoring. Multiple studies have applied DL techniques to improve predictive performance. For example, a neural network model was proficiently employed to assess flood hazard risk distribution within the Tajan watershed, utilizing a spectrum of hydrological, topographical, environmental, and anthropogenic factors [10]. The Deep Neural Networks (DNNs) and multi-criteria decision analysis (MCDA) are combined to develop the flood risk map [11] [12]. A convolutional neural network (CNN) is also applied for flood hazard mapping [13]. The existing ML-based and DL-based models show the ability of these models to capture complex and non-linear interactions among various features in predicting flood inundation and risk. However, the main focus in the existing machine learning-based models is to predict flood inundation (exposure) with limited attention to property flood risk, leaving two important research questions unanswered:

(1) To what extent do various intertwined hydrological, topographic, and built-environment features explain variations in the property flood risk disposition of different areas of cities? and (2) Are the underlying features and their interactions shaping property flood risk predisposition universal across different cities? Answering these questions is essential for decoding the flood risk genome of cities for understanding the inherent predisposition of different spatial areas and informing urban design strategies for alleviating flood risk disposition in different areas of cities.

Recognizing this important gap, this study presents FloodGenome (Figure 1) which is an explainable machine learning model trained and tested to specify property flood risk propensity of spatial areas based on various hydrologic, topographic, and built-environment features. This novel approach to flood risk assessment by integrating a random forest model with k-means clustering, utilizing extensive flood damage claims data from the U.S. National Flood Insurance Program (NFIP). Covering the period from 2003 through 2023 and focusing on four metropolitan statistical areas (MSAs), this research aims to develop a predictive model that not only assesses flood risk with high accuracy but also identifies key determinants of flood susceptibility across different urban settings. By analyzing building damage ratios, flood claim counts, and incorporating various hydrological, topographical, and built-environment features, the study categorizes census block groups (CBGs) into distinct flood risk levels. Furthermore, this research examines the transferability of the developed model across different MSAs, evaluating the consistency of factors influencing flood risk and the potential for a unified, scalable approach to flood risk assessment. This endeavor not only contributes to the refinement of flood prediction models but also provides valuable insights for urban planning and disaster management strategies, aiming to mitigate the adverse impacts of flooding and enhance community resilience.

In sum, the main contributions of this study are five-fold: first, the introduction of the concept of flood risk predisposition as an event-independent measure of the extent to which urban areas are susceptible to property damage if exposed to weather hazards. This new concept provides a new measure for flood risk rating for evaluating the spatial profile of property flood risk. Evaluating the flood risk disposition of spatial areas irrespective of weather events can inform urban plans and flood risk reduction measures to alleviate the flood propensity of areas. Second, by leveraging machine intelligence, the FloodGenome model and its performance reveal the relative importance of hydrological, topographic, and built-environment features and their interactions that shape property flood risk predisposition of spatial areas. Analogically, as genetic features influence predisposition to certain diseases, the identified hydrological, topographic, and built-environment features shape property flood risk disposition. Third, the results of the model transferability evaluation unveil that the intertwined features are universal across different cities. This universality of the underlying features and their interaction enables the transferability of strategies and policies for urban planning and flood risk management across different cities and regions. Fourth, the FloodGenome model enables examination of the extent to which urban development patterns change property flood risk disposition across different areas of cities. This approach provides a shift in focus from event-based flood risk evaluation to event-independent flood risk predisposition assessment. The foresight is essential to proactively identify areas whose development should be controlled to avoid exacerbating their flood risk predisposition. Fifth, after being trained at coarser spatial scales, the FloodGenome model can be used for specifying property flood risk predisposition at finer spatial resolutions to provide a flood risk rating tool and insights for flood risk management plans and policies supplementing existing flood plain maps. The study also contributes to the ongoing efforts for harnessing machine intelligence for advancing the risk and resilience characterization of cities.

These outcomes provide interdisciplinary researchers in urban planning, disaster science, geography, and engineering with new models and insights to better characterize, understand, and analyze the underlying features that shape property flood risk predisposition of urban areas. The outcomes also provide flood managers, emergency managers, and public officials with new tools and knowledge for alleviating flood risk in cities through integrated urban design strategies. For example, FloodGenome can inform urban zoning policies to limit development in low-lying areas with low height above the nearest drainage, which is among the most important features identified in this study. Also, flood managers can use flood genome to evaluate strategies for flood risk reduction in different areas based on the main features of the areas. These outcomes would complement the existing flood risk management methods and move us closer to better management of flood risk in cities.

## 2 Method

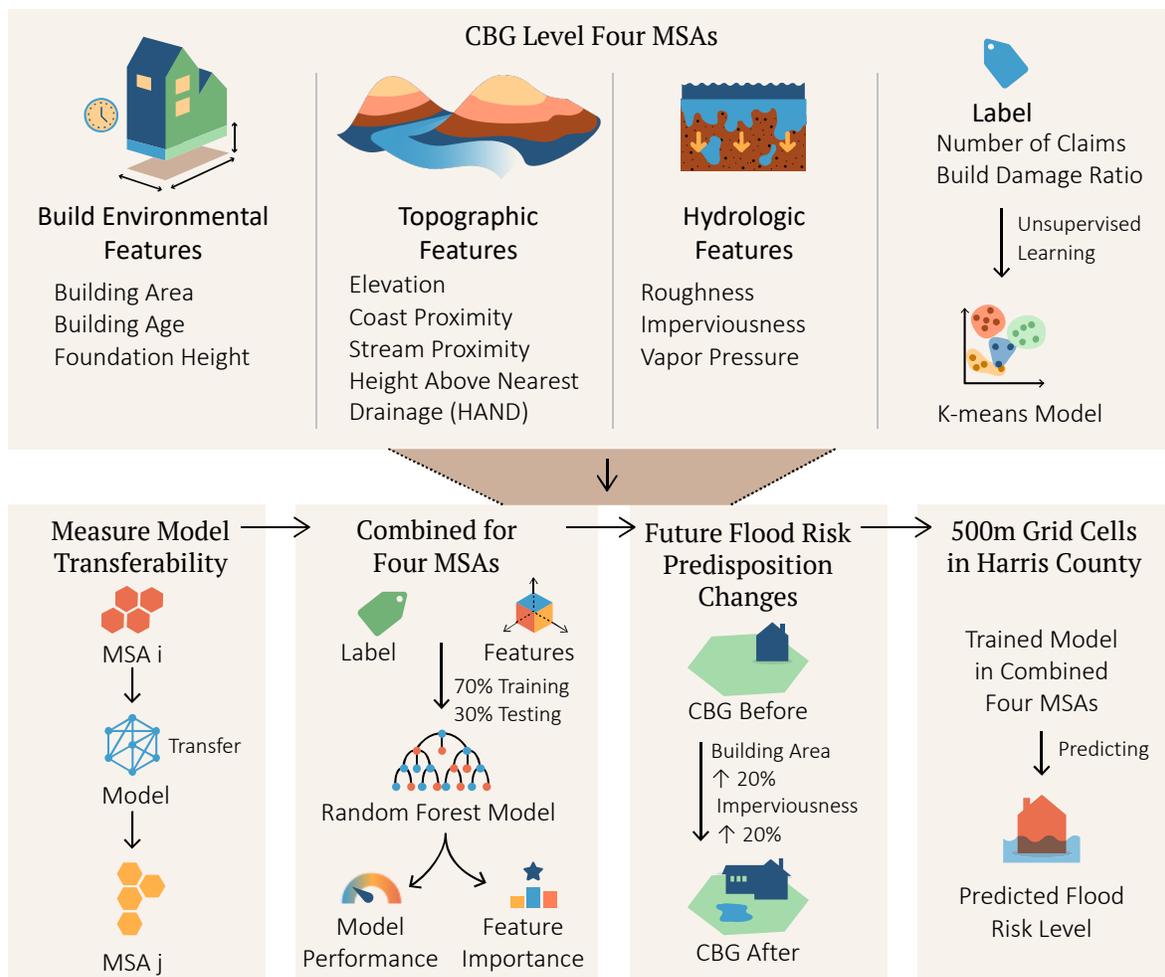

**Figure1. Illustration of the methodological framework.** Evolving the integration of built environmental, topographic, and hydrologic characteristics alongside k-means clustering to delineate distinct flood risk groups. Initially, random forest models are individually trained for each metropolitan statistical area. Following this, First, a rigorous assessment of model transferability ensues, ensuring robustness across MSAs through stringent similarity confirmation procedures. Second, features and labels from four MSAs undergo amalgamation, prompting model retraining and subsequent evaluation of model performance. This evaluation is accomplished through the computation of the ROC-AUC scores, complemented by feature importance elucidation via SHAP analysis. Third, flood risk predisposition changes are measured by increasing building area and imperviousness by 20%. Ultimately, this process culminates in the prediction of flood risk levels within a 500m grid cell resolution specifically tailored to Harris County, Texas.

## 2.1 Data description and processing

### 2.1.1 NFIP flood claim data

In this study, we analyzed flood damage claims sourced from the U.S. National Flood Insurance Program [14] over the period spanning 2003 through 2023, focusing on four metropolitan statistical areas: Houston, Miami, New Orleans, and New York. Table 1 provides a breakdown of the number of claims and the number of CBGs within each respective MSA. The dataset about flood claims encompasses information on building damage amounts, building property values, and the respective dates of loss.

To facilitate our analysis, The building damage ratio for each structure was calculated by the following formula:

$$Building\ damage\ ratio_i = \frac{building\ damage\ amount_i}{builidng\ property\ value_i} \qquad (1)$$

Here, $i$ stands for each claimed building in NFIP flood claims data. Given that these buildings incurred damage across different years, we adjusted the building damage amounts and building property values to 2023 levels, factoring in inflation rates before computing the building damage ratios. Subsequently, we computed the mean value of these ratios for each census block group within the various MSAs. Concurrently, we also determined the total number of damaged buildings within each CBG.

**Table 1. Summary of NFIP flood claims and census block groups in select MSAs**

| MSA | Number of CBGs | Number of flood claims since 2003 |
| --- | --- | --- |
| **Houston** | 3,021 | 130,460 |
| **Miami** | 3,420 | 19,286 |
| **New Orleans** | 1,110 | 156,795 |
| **New York** | 14,376 | 172,287 |

After obtaining the mean value of the building damage ratio and the total number of damaged buildings, we applied the K-means clustering algorithm to these two features to categorize the CBGs into distinct flood risk groups. The determination of the number of groups for each metropolitan statistical area was based on the analysis of an elbow plot generated by the K-means model. Subsequently, we assessed the flood risk levels of these groups by calculating a flood risk value using the following formula:

$$Flood\ risk\ value_i = E(the\ number\ of\ damaged\ buildings)_i * E(building\ damage\ ratio)_i \quad (2)$$

Here, *i* is the number of clusters determined by the K-means model. It is important to note that a higher flood risk value corresponds to a greater level of flood risk.

Based on the elbow graph results, we classified flood risk into four levels. CBGs without flood claims data were grouped into level 0. The number of CBGs in each level for four MSAs is shown in Table 2.

**Table 2. Initial distribution of census block groups by flood risk levels in four MSAs**

| MSA | Level 0 | Level 1 | Level 2 | Level 3 | Level 4 | Highest imbalance ratio |
|---|---|---|---|---|---|---|
| Houston | 282 | 1720 | 833 | 181 | 5 | 0.003 |
| Miami | 1134 | 199 | 1915 | 156 | 16 | 0.008 |
| New Orleans | 14 | 602 | 304 | 165 | 25 | 0.023 |
| New York | 9301 | 3836 | 108 | 946 | 185 | 0.012 |

Typically, the high class imbalance problem is defined as a majority-to-minority class ratio between 100:1 and 10,000:1 [15]. From Table 2, we found that there exists a high class imbalance problem in some MSAs. To address the extreme imbalance issue, we combined the minority cluster with the nearest cluster. For example, for the Houston MSA and the Miami MSA, we combined data in level 4 with level 3 to form a revised level 3. To make the number of clusters consistent with each other. We also made some changes in the New Orleans MSA and New York MSA. For the New Orleans MSA, we combined the data in level 0 with level 1 to form a revised level 1. For the New York MSA, we combined the data in level 2 with level 1 to form a revised level 1. The updated number of CBGs in each level for the four MSAs is shown in Table 3. In the following context, we defined level 0 as low flood risk, level 1 as medium flood risk, level 2 as high flood risk, and level 3 as extreme flood risk. The distribution of flood risk level by k-means method across four MSAs is shown in Figure 2.

**Table 3. Revised distribution of census block groups by combined flood risk levels in four MSAs**

|  | Flood risk level | | | |
| --- | --- | --- | --- | --- |
| MSA | Low | Medium | High | Extreme |
| Houston | 282 | 1720 | 833 | 186 |
| Miami | 1134 | 199 | 1915 | 172 |
| New Orleans | 616 | 304 | 165 | 25 |
| New York | 9301 | 3944 | 946 | 185 |
| Total | 11333 | 6167 | 3859 | 568 |

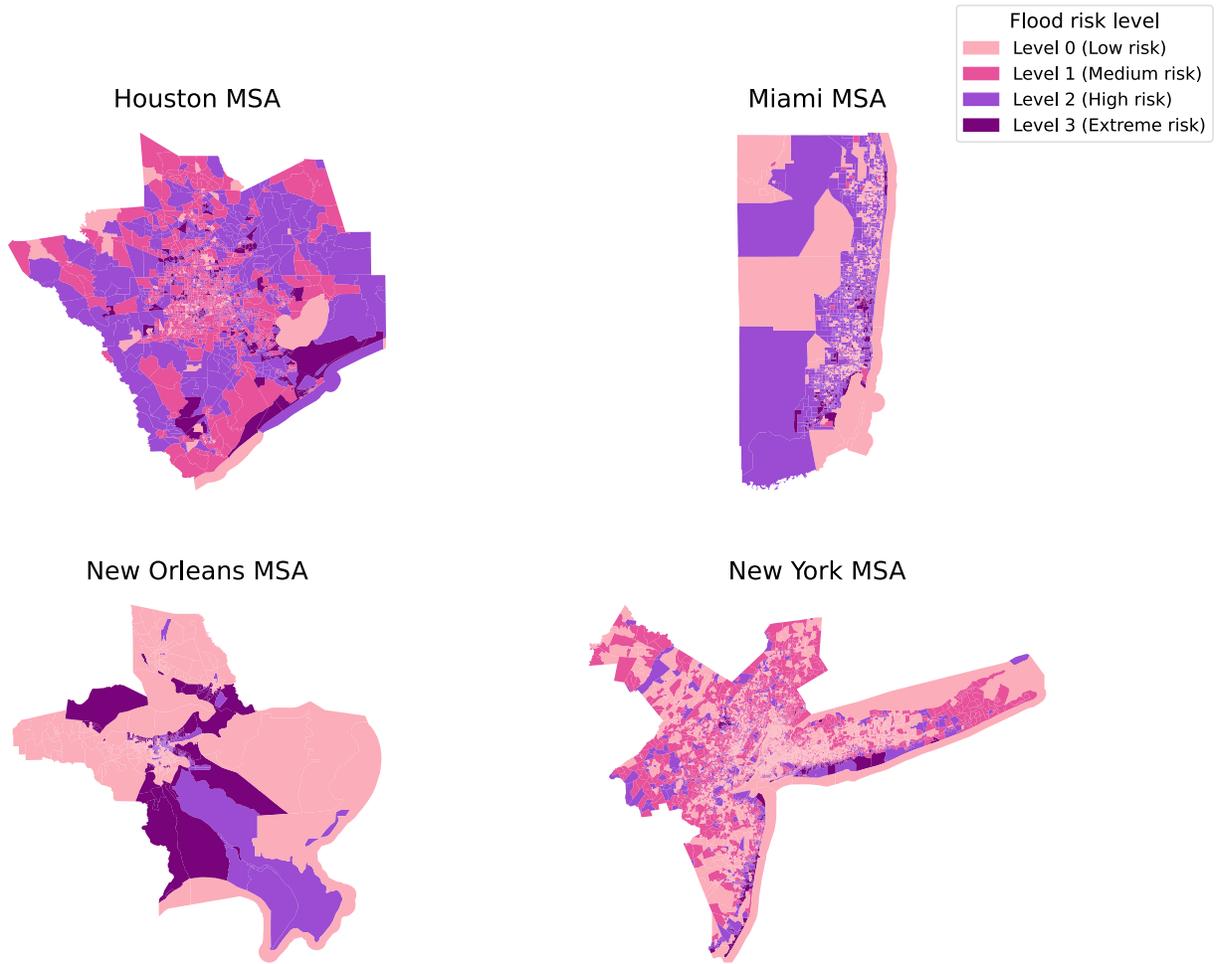

**Figure 2. Flood risk level across different CBGs for four MSAs obtained from k-means clustering**

2.1.2 Hydrological features

**Imperviousness:** Impervious surfaces comprise artificial structures, such as pavements and buildings, which are constructed using water-resistant materials. These surfaces effectively create a barrier that prevents rainwater from infiltrating the soil and impeding the natural recharge of groundwater [16] [17]. Consequently, they contribute to the occurrence of urban flooding. The data regarding imperviousness was sourced from the National Land Cover Database for the year 2019 [18]. The original dataset quantifies urban impervious surfaces as a percentage of the developed surface within each 30-meter pixel across the United States.

**Height above nearest drainage:** Height above the nearest drainage (HAND) calculates the vertical distance between the location and its nearest stream [19]. It is often used to assess the flood risk of a particular area [20] [21]. Areas with lower HAND values are closer to drainage features and are more likely to flood during heavy rainfall or other flood-inducing events. The HAND dataset was downloaded from the University of Texas National Flood Interoperability Experiment continental flood inundation mapping system at a 10-meter resolution [22].

**Roughness:** Surface roughness, within the context of hydrology, pertains to the terrain's capacity to act as a momentum dissipator in the presence of overland water flow [23]. The role of roughness is of considerable significance in numerous hydrological models [24]. In this research, Manning's coefficient is employed as a representative parameter for expressing surface roughness. The computation of Manning's coefficient is grounded on land cover data sourced from the National Land Cover Database [25]. Subsequently, surface roughness values are computed for each specific land cover type, following the methodology proposed by Kalyanapu et al. [10].

**Vapor pressure:** Vapor pressure manifests as the presence of vapor above the surface of a liquid. Elevated vapor pressure levels have the potential to cause increased atmospheric humidity. It is worth noting that heightened vapor pressure can be a contributing factor to the occurrence of substantial rainfall [26] [27]. In this study, we acquired annual average vapor pressure data from the NASA Earth Daymet dataset, which is characterized by a spatial resolution of 1 kilometer [28].

Given the scale of our study, which operates at the census block group level, we computed the mean imperviousness value for CBGs to serve as our imperviousness feature.

### 2.1.3 Topographic features

**Elevation:** While the HAND feature measures the vertical distance between the location and the nearest stream, elevation is used to evaluate the location with sea level. It can distinguish the high and low areas compared with the same condition, which is sea level. Elevation is frequently used to model flood hazards [29] [30] [31] [32]. Low-lying areas adjacent to rivers or coastal regions are at greater risk of experiencing severe, long-lasting floods, while higher elevations are more likely to see limited or less severe flooding during the same event. Elevation data was collected from the elevation package using Python in a 30-meter resolution [33].

**Distance to stream and coast:** The proximity to streams and coastlines has been established as a statistically significant indicator in the prediction of flood damage [34]. The determination of distances from specific CBGs to both streams and coastlines involves the use of data extracted from the National Hydrography Dataset, which provides comprehensive information on the geometry of streams and coastline features. This data served as the foundation for our calculations. Initially, we filtered the streams and coasts data in specific MSAs to reduce computing time. Subsequently, we identified the nearest point on the stream and coastline concerning the centroid of each CBG. Using these reference points, we proceeded to compute the distances between the CBGs and the respective nearest stream and coastline locations. Consequently, for each CBG within distinct metropolitan statistical areas, two lists were generated: one denoting the distances to streams and the other indicating the distances to coastlines. The ultimate selection for the distance-to-stream and distance-to-coast features was determined by identifying the minimum value from these two lists.

2.1.4  Built-environment features

**Building age:** Building age data describing the median year structures within the census block were built, and were downloaded from the National Structure Inventory dataset [35].

**Building area:** Building area data, the estimated square footage of the structures, were downloaded from the NSI dataset [35].

**Foundation height:** Foundation height data, the height of the structure foundation measured from ground elevation, are downloaded from the NSI dataset [35].

## 2.2  Methods

2.2.1  K-means

The k-means clustering [36] endeavors to allocate n observations into k clusters, wherein each observation is assigned to the cluster with the nearest mean. The determination of the optimal number of clusters is facilitated by the use of the elbow method, which relies on the mean value of the building damage ratio and the total count of damaged buildings. The elbow graph portrays the within-cluster-sum-of-square values on the y-axis corresponding to various values of K—the number of clusters—delineated on the x-axis. The elbow graphs for four distinct MSAs are presented in the supplementary information section. Subsequent to the analysis of these graphs, flood risk is categorized into four levels across all four MSAs, informed by the insights gleaned from the elbow graphs.

2.2.2  Random forest classifier

The random forest classifier model [37], a supervised learning algorithm, is an ensemble learning technique used for tasks such as classification and regression. This method generates numerous decision trees during training then determines the output class based on either the mode of the classes or the mean prediction of the individual trees. A random forest comprises a collection of independent decision trees. The selection of the random forest model for classification was motivated by its ability to address imbalanced data through the assignment of weights to different clusters. These weights play a pivotal role in influencing the splitting criterion during the construction of decision trees within the random forest algorithm, thereby assigning greater significance to samples from minority classes.

2.2.3  SMOTE

Aside from adjusting the class weight in the random forest classification model to alleviate the imbalanced data problem, we also applied the synthetic minority oversampling technique (SMOTE) method with random forest to check the performance of the random forest. The SMOTE method [38] is an upsampling method widely used in dealing with imbalanced datasets [39] [40] [41] . The SMOTE method will add artificially simulated new samples to the minority class to alleviate the severe class imbalance in the original data. First, it will randomly choose a sample belonging to the minority class. Second, it will choose k-nearest neighbors around it. Third, it will randomly choose a sample from the k-nearest neighbors. Fourth, it will randomly choose a point between the minority point and the chosen nearest neighbor.

2.2.4 Evaluation matrix

The receiver operating characteristic area under the curve (ROC-AUC) is used to evaluate the model performance in this research. The ROC curve plots the true positive rate against the false positive rate. The true positive rate is the ratio of positive instances correctly identified as positive out of all actual positives. The false positive rate is the ratio of negative instances incorrectly identified as positive out of all actual negatives. AUC and ROC are widely used in binary classification problems; in this study we extended it to multi-class classification scenarios using strategies like one-vs-rest. For each class, a separate binary classifier is trained against all the other classes combined. The higher the score, the better the model can distinguish the correct class.

## 2.3 Measurement of feature importance

This study quantifies the contribution of each feature variable to flood risk level using the Shapley value, a concept from game theory [42], as a unified measure of feature importance. In the context of machine learning models, SHAP values attribute the effect of each to the prediction made by the model for any given sample.

# 3 Results

## 3.1 Model performance and transferability

3.1.1 Model performance

Table 4 presents the area under the receiver operating characteristic curve (AUC) scores derived from a random forest model subjected to fine-tuning through five-fold cross-validation, employing a training-test partition of 70% and 30%. This validation methodology ensures a robust evaluation of the model's predictive performance across various metropolitan statistical areas. The table juxtaposes the effectiveness of the SMOTE against adjustments in hyperparameter class weights for addressing dataset imbalances. Notably, the adjustment of class weights appears to offer better results compared to the implementation of the SMOTE method in mitigating the effects of imbalanced data in this study. This observation may be attributed to the synthetic generation of samples within the minority class via SMOTE, potentially exacerbating issues related to overfitting, as opposed to the augmentation of existing samples through class weight adjustments [43].

**Table 4. Comparative AUC scores for random forest model using SMOTE and class weight adjustments across four MSAs**

| MSA | SMOTE | Class weight |
|---|---|---|
| **Houston** | 0.78 | 0.79 |
| **Miami** | 0.76 | 0.77 |
| **New Orleans** | 0.92 | 0.93 |
| **New York** | 0.87 | 0.88 |

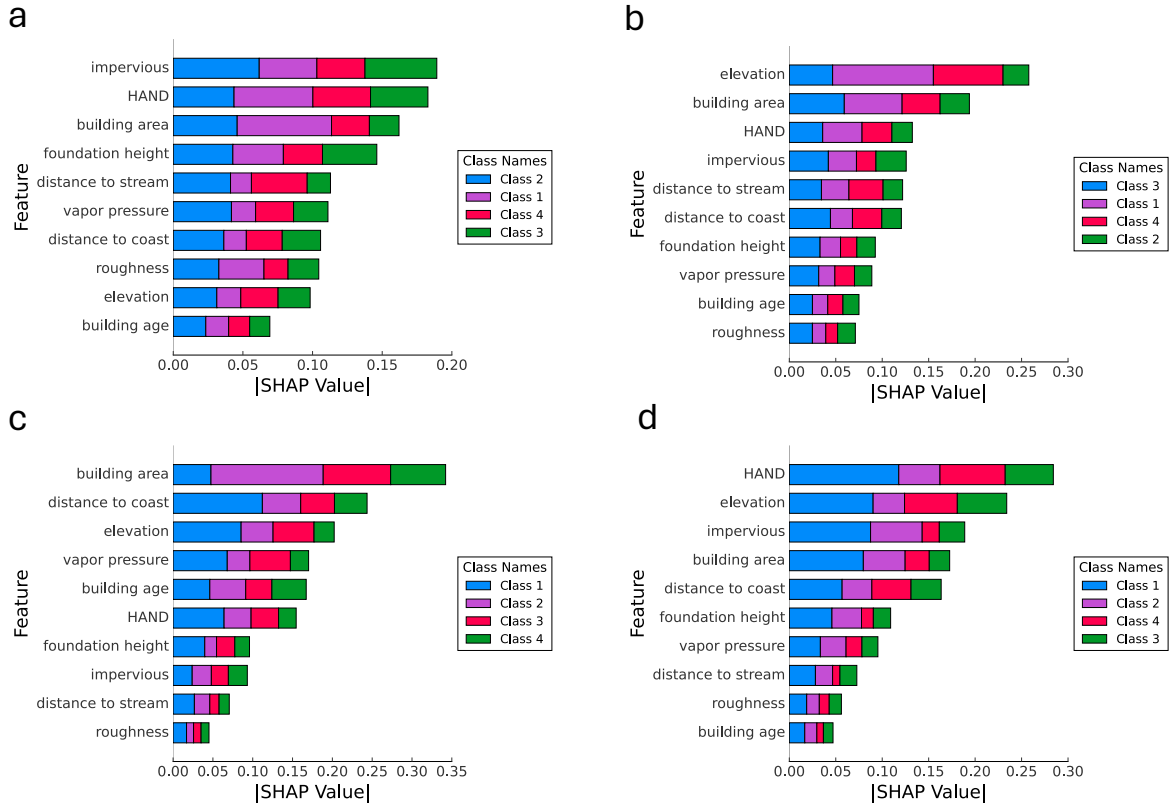

**Figure 3. SHAP value summary plot for four MSAs:** (a) Houston (b) Miami (c) New Orleans (d) New York. The y-axis shows the flood risk disposition features. The x-axis represents the sum of absolute SHAP values for each feature. Colors represent distinct clusters with feature importance ranked from highest to lowest.

Figure 3 shows the ranking of feature importance across four MSAs. Based on the scale of different absolute SHAP values shown in different MSAs, we examined the top five features across the four MSAs. Building area is consistently ranked as one of the top five features (plots (a), (b), (c), and (d)). This result indicates that regardless of the hydrologic and topographic features, the extent of building development (i.e., exposure) is a significant predictor in flood risk predisposition outcomes. This result highlights the importance of exposure control measures (e.g., controlled development in flood-prone areas) to manage property flood risk. HAND and Impervious, appear as important features in subplots (a), (b), and (d). These features are influenced by land development flood control and management measures. Their prominence in shaping property flood risk suggests that for urban development patterns, increasing imperviousness and altering HAND would exacerbate flood risk. Elevation is another feature appearing in subplots (b), (c), and (d), and suggesting the importance of reducing exposure in low-lying areas for alleviating flood risk extent. The commonality of different features among the top five features across the four MSAs suggests that the underlying features and their interactions shaping flood risk predisposition might be universal across different cities. To further evaluate this universality of the urban flood risk genome, we examined model transferability performance.

### 3.1.2 Model transferability across cities

In the next step, we examined the universality of the urban flood risk genome based on model transferability performance analysis. If the underlying features and their interactions with flood risk are universal across MSAs, then a model trained on data from one MSA could yield accurate predictions of the risk level of CBGs when applied to other MSAs. The model transferability analysis is based on the evaluation of model performance, measured by the ROC-AUC score, utilizing the model trained on one MSA to predict flood risks in three other MSAs. The findings regarding model transferability are presented in Table 5, where each entry corresponds to the ROC-AUC score attained when the model trained on data from the MSA specified in the first column is tested on data from the MSA indicated at the top of the columns. Notably, most transferred models exhibit good performance compared to the original model. Based on the results in Table 5, we can conclude that models trained on data from one MSA can yield accurate predictions when applied to others. These results suggest that the underlying intertwined features influencing flood risk could be universal across different cities and regions. Also, this result supports the idea of combining the datasets to train a new, comprehensive model (i.e., FloodGenome) for property flood risk classification across areas of different cities. We present the result of the model trained and tested on the combined data of the four MSAs in the following section.

**Table 5. Cross-MSA flood risk prediction transferability scores on testing data**

|  | Houston | Miami | New Orleans | New York |
|---|---|---|---|---|
| **Houston** | 0.787 | 0.786 | 0.779 | 0.785 |
| **Miami** | 0.773 | 0.775 | 0.777 | 0.776 |
| **New Orleans** | 0.929 | 0.929 | 0.932 | 0.932 |
| **New York** | 0.881 | 0.881 | 0.880 | 0.882 |

## 3.2 Model performance and feature importance for combined MSA data

The FloodGenome model trained and tested on the combined data showed very good performance. Figure 4 presents the receiver operating characteristic curves for the four classes of predictions. These curves plot the true positive rate (TPR) against the false positive rate (FPR) at various threshold settings, providing a visual representation of the trade-off between sensitivity and specificity. Each individual lines represent the respective flood risk levels (low to extreme). The closer a curve follows the left-hand border and then the top border of the ROC space, the more accurate the test. Thus, a curve near the top-left corner indicates an excellent level of discrimination. The diagonal dashed line represents a no-skill classifier; a classifier that cannot distinguish between classes will have a ROC curve that lies along this line.

Based on the results shown in Table 6 and Figure 4, the analysis indicates that the FloodGenome model, when adjusted for class weights to address dataset imbalances, exhibits varied effectiveness across different flood risk levels. The ROC-AUC scores presented in Table 6—0.89 for both level 0 and level 2, 0.84 for level 1, and 0.96 for level 3—highlight the model's particular strong performance in identifying extreme flood risk areas (level 3). This result suggests the model is most effective at distinguishing the highest risk areas, potentially offering valuable insights for specifying areas with the greatest flood risk.

**Table 6. AUC Scores for random forest model using class weight adjustments**

|                | Level 0 | Level 1 | Level 2 | Level 3 |
|----------------|---------|---------|---------|---------|
| **ROC-AUC score** | 0.89    | 0.84    | 0.89    | 0.96    |

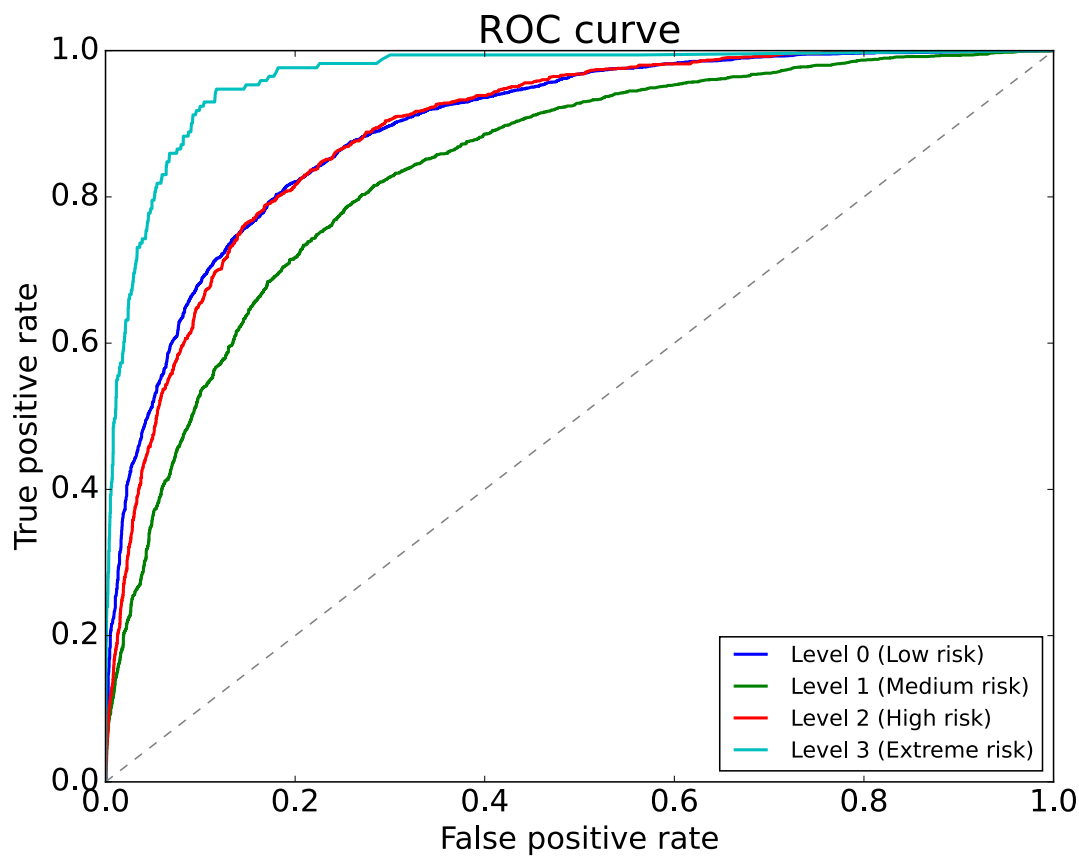

**Figure 4: Receiver operating characteristic curves.** Each curve represents the trade-off between the true positive rate and the false positive rate for a given flood risk level.

### 3.3 Universal features shaping property flood risk disposition

The hierarchical ranking of features delineates a continuum ranging from direct, broadly applicable contributors to flood risk levels to localized, indirect determinants that shape flood risk and property damage outcomes. Figure 5 shows the SHAP summary plot which can reveal the importance of features in the FloodGenome model trained on the combined datasets of the four MSAs. The most important features, arranged in descending order of importance, include vapor pressure, height above nearest drainage, elevation, impervious surfaces, building area, distance to coast, foundation height, distance to stream, building age, and roughness. Each feature's relative impact on the model's output is visually represented by bar lengths, with color segments delineating the distribution of impact across distinct risk levels. This result unveils the universal features shaping property flood risk levels in cities. Among the top five most top features, vapor pressure is the most important, as it captures precipitation levels. Elevated vapor pressure correlates with heightened precipitation frequency and intensity, directly contributing to escalated flood risk levels. HAND, by delineating the elevation differential between a point and the nearest water body, the second most important feature, captures potential stormwater accumulation and flow trajectories, thereby emerging as a pivotal predictor of flood risk levels. Elevation, similar to HAND, represents a fundamental hydrological parameter governing water flow dynamics. Low-lying areas have elevated flood risk levels. Furthermore, impervious surfaces and building areas underscore the urbanization-induced features escalating flood risk levels, as impervious surfaces impede water absorption, increasing runoff, and thereby exacerbating urban flooding risk, especially in densely populated locales. Building area captures the extent of flood exposure, one of the primary elements contributing to flood risk levels. The hierarchical ranking of features thus delineates a continuum ranging from direct, broadly applicable contributors to flood risk levels to localized, indirect determinants that shape flood risk and property damage outcomes.

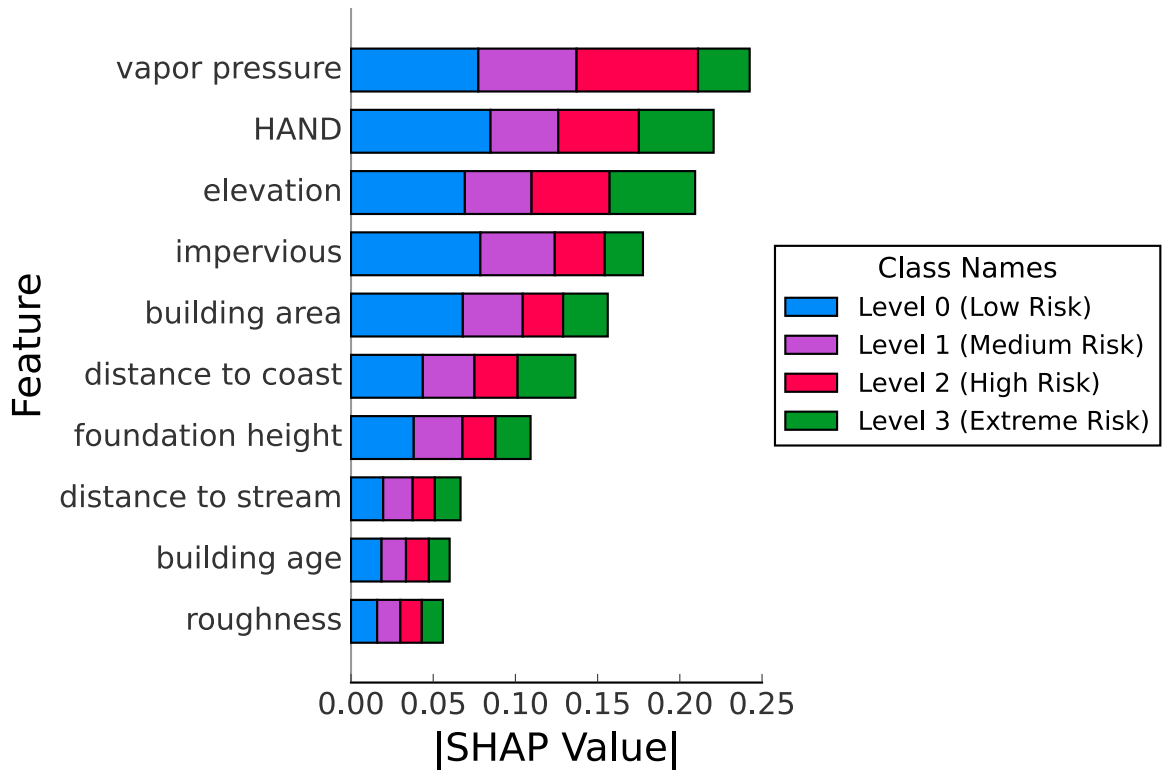

**Figure 5. SHAP value summary for combined data for four MSAs.** The y-axis shows flood risk disposition feature names. The x-axis represents the sum of absolute SHAP value for each feature. Colors represent distinct clusters with feature importance ranked from highest to lowest.

### 3.4 Future urban development and flood risk predisposition changes

In the next step, using the FloodGenome model, we examine the influence of urban development features (impervious and building area) on changes in property flood risk disposition of different areas across the four MSAs. Urbanization often leads to an increase in impervious surfaces, such as roads, sidewalks, and buildings, which can exacerbate the risks and consequences of flooding by reducing the land's natural absorption capacity. The experiment presented in this part seeks to provide a future outlook regarding the extent to which a 20% increase in impervious surface area or a 20% increase in building area might elevate flood risk levels across different CBGs.

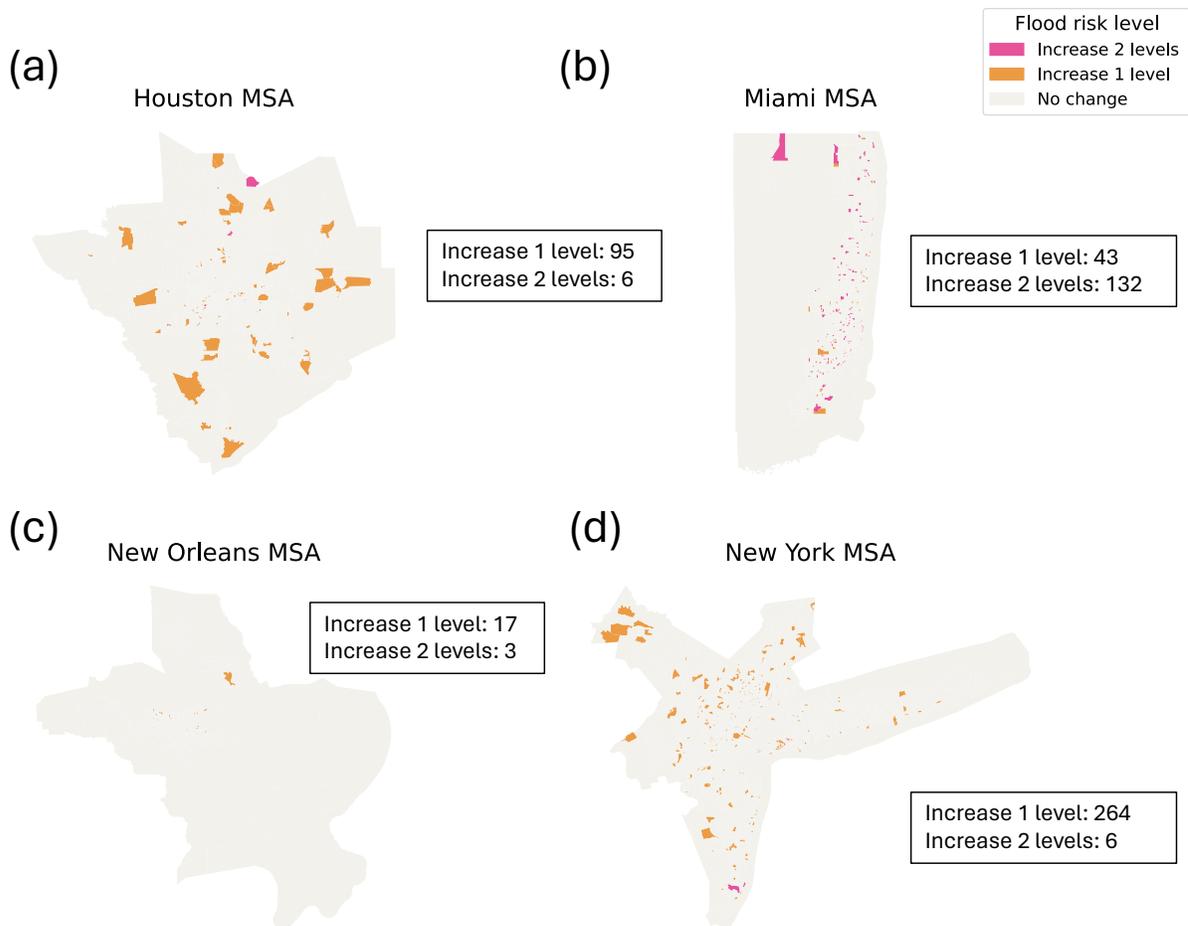

**Figure 6. Projected increase in flood risk due to urban impervious surface expansion across four MSAs.** (a) Houston (b) Miami (c) New Orleans (d) New York. Pink indicates the census block groups with an anticipated increase in flood risk by two levels. Yellow indicates an increase by one level. Gray denotes areas with no expected change. The number for each subplot shows the number of census block groups increase 1 level and 2 levels.

The visualized results from the FloodGenome model for the experiment of increasing imperviousness by 20%, are shown in the Figure 6. (A similar plot related to increasing building area by 20% is shown in SF.2). The plot, highlighting areas within the Houston, Miami, New Orleans, and New York MSAs, differentiates between CBGs that are predicted to experience a one-level increase, a two-level increase, or no change in flood risk. Areas marked with darker shades indicate a greater anticipated change in flood risk level, implying the need for controlling development (or implementing low-impact development). The nuanced understanding gained from these results can directly inform future urban development plans to identify areas in which further development should be strictly controlled. The results can also inform strategic zoning laws that could mitigate the impact of increased impervious surfaces. These results can supplement flood maps to inform integrated urban design strategies that proactively identify the effects of development on the flood risk of areas and identify strategies to mitigate flood risk escalation in different areas of cities.

### 3.5 Specifying flood risk disposition at finer resolution

Given the enhanced predictive capability of the FloodGenome model in assessing flood risk levels utilizing census block group data, an intriguing prospect arises when considering the delineation of flood risk levels for smaller geographic areas. Motivated by this idea, we have undertaken the task of generating high-resolution flood risk levels at a spatial scale of 500m x 500m, as depicted in Figure 7 (example of fine resolution flood risk rating for Harris County). Using the FloodGenome model trained and tested on data at the CBG level, we computed the input features at the finer spatial resolution and specified the flood risk levels of smaller spatial areas. The specification of flood risk levels of spatial areas at a finer resolution provides a new tool and insights to supplement flood maps for examining the flood risk profile of communities at finer spatial resolution. The finer resolution results provide a flood risk rating tool and insights to supplement existing flood plain maps to inform flood risk management plans and policies. The main utility of the FloodGenome model in producing fine-resolution flood risk rating is its ease of use to reflect changes in urban development and land changes for evaluating the changes in flood risk levels as shown in the earlier results for the CBG-level analysis.

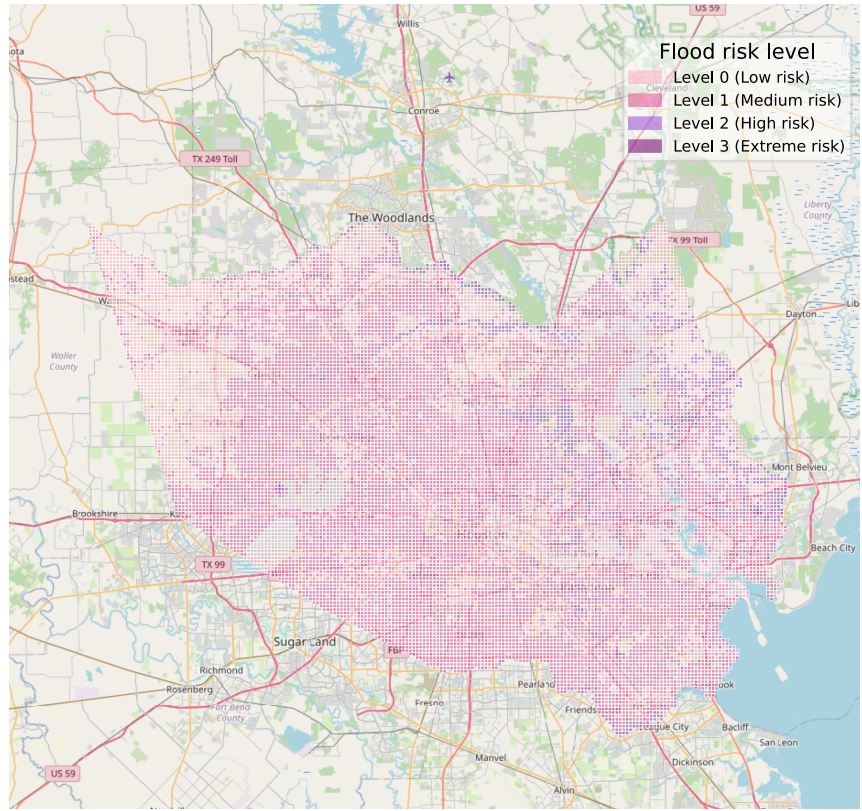

**Figure 7. Flood risk distribution map of Harris County at 500m resolution**

## 4 Concluding Remarks

This research introduces FloodGenome, an interpretable machine learning model designed to assess the property flood risk predisposition of specific areas based on the underlying intertwined hydrological, topographical, and constructed environment characteristics. The concept of property flood risk predisposition is introduced as an event-independent way to evaluate the extent of susceptibility to property damages in spatial areas of cities. The study leverages flood damage claims data from the U.S. National Flood Insurance Program spanning 2003 through 2023 across four metropolitan statistical areas, employing k-means clustering to segregate census block groups into four unique flood risk predisposition categories based on building damage ratios and flood claim frequencies. Subsequently, random forest models were developed and evaluated to determine CBGs' flood risk levels utilizing a range of hydrological, topographical, and built environment attributes, demonstrating that these features significantly account for the variations in flood risk susceptibility of spatial areas. The research explored the adaptability of these models, showing their capability to predict flood risk levels in CBGs of different cities, and highlighting the consistency of feature significance and interactions in influencing flood risk across varied locales. In addition, the study assesses the impact of urban development factors, such as impervious surfaces and building volume, on escalating flood risk in cities, pinpointing regions where intensified development could heighten flood dangers, thereby guiding urban planning and policy making. Finally, by applying FloodGenome at a more detailed spatial resolution (500 m x 500 m grid cells), the study offers nuanced insights into the flood risk disposition of areas, enhancing flood exposure maps to support community flood prevention and risk mitigation strategies.

The study outcomes involve multiple scientific and practical contributions. First, the introduced concept of inherent property flood risk disposition provides a hazard-independent way to evaluate the extent to which a spatial area is susceptible to property damage when exposed to weather hazards. This concept provides new perspectives for evaluating flood risk of urban areas complementing the existing prominent approaches focusing on projecting weather hazards in predicting flood inundation exposure. Flood researchers and flood managers in communities can use the rating of spatial areas in terms of their flood risk predisposition along with flood maps to better evaluate flood susceptibility and exposure across different CBGs.

Second, we leverage machine intelligence to create the FloodGenome model. The performance of the model reveals the relative importance of hydrological, topographic, and built-environment features. The interactions of these features effectively explain variations in property flood risk predisposition of spatial areas. Unlike existing physics-based models that provide limited interpretability, our explainable ML model unravels the underlying features and their non-linear interactions shaping city flood risk profiles. Specifying the relative importance of the underlying features of flood risk predisposition and their interactions is critical for informing flood mitigative urban development. Analogically, awareness of genetic predisposition to disease is important for taking preventive health measures, just as the knowledge of relative importance and interaction among hydrological, topographic, and built-environment features is essential for mitigative urban development. These insights can inform urban development and growth plans and policies. For example, the insights can guided urban zoning policies to limit development in low-lying areas with low height above the nearest drainage which are among the most important features identified in this study.

Third, the results of the model transferability evaluation unveil that the intertwined features are universal across different cities. This universality of the underlying features and their interaction enable transferability of strategies and policies for urban planning and flood risk management across different cities and regions. The uncovering of the universality of the urban flood risk genome is particularly noteworthy for transferring effective flood risk management strategies across different locales. This is analogous to using a proven preventative measure that has proven effective in one individual having a genetic predisposition to a certain disease on others with the same genetic marker. This outcome encourages community flood managers and city officials to adopt effective flood risk reduction strategies from other areas.

Fourth, the FloodGenome model enables examining the extent to which urban development patterns would change property flood risk disposition across different areas of cities. This approach provides a shift in focus from hazard-focused flood risk evaluation to hazard-independent flood risk predisposition assessment. This foresight is essential to proactively identify areas whose development should be controlled to avoid exacerbating their flood risk predisposition. Also, unlike effort- and computationally-expensive physics-based models, FloodGenome enables efficient evaluation of how changes in features such as imperviousness and building area due to development would exacerbate flood risk predisposition in different areas of a city. This foresight are crucial for urban scientists and urban planners to proactively evaluate the effect of urban development and growth patterns on property flood risk of areas. Fifth, after being trained at coarser spatial scales, the FloodGenome model can be used for specifying property flood risk predisposition at finer spatial resolutions to provide a flood risk rating tool and insights for flood risk management plans and policies supplementing existing flood plain maps.

These outcomes provide interdisciplinary researchers in urban planning, disaster science, geography, and engineering with new models and insights to better characterize, understand, and analyze the underlying features that shape property flood risk predisposition of urban areas. The outcomes also provide flood managers, emergency managers, and public officials with new tools and knowledge for alleviating flood risk in cities through integrated urban design strategies. For example, FloodGenome can inform urban zoning policies to limit development in low-lying areas with low height above the nearest drainage which are among the most important flooding predictors identified in this study. Also, flood managers can use flood genome to evaluate strategies for flood risk reduction in different areas based on the main features of the areas. These outcomes would complement the existing flood risk management methods and move us closer to better management of flood risk in cities.

## Acknowledgements

This material is based in part upon work supported by the National Science Foundation under CRISP 2.0 Type 2 No. 1832662 grant. Any opinions, findings, conclusions, or recommendations expressed in this material are those of the authors and do not necessarily reflect the views of the National Science Foundation.

## Code availability

The code that supports the findings of this study is available from the corresponding author upon request.

# Supplementary Information

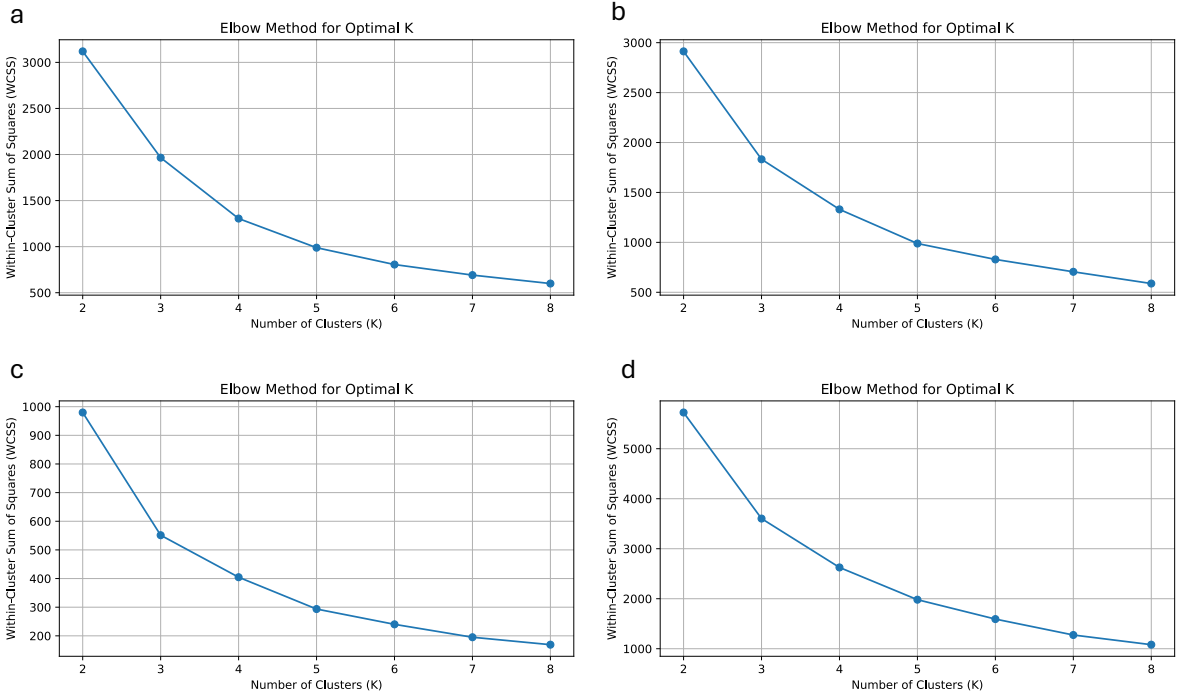

**SF 1. Elbow graph for** (a) Houston MSA (b) Miami MSA (c) New Orleans MSA (d) New York MSA

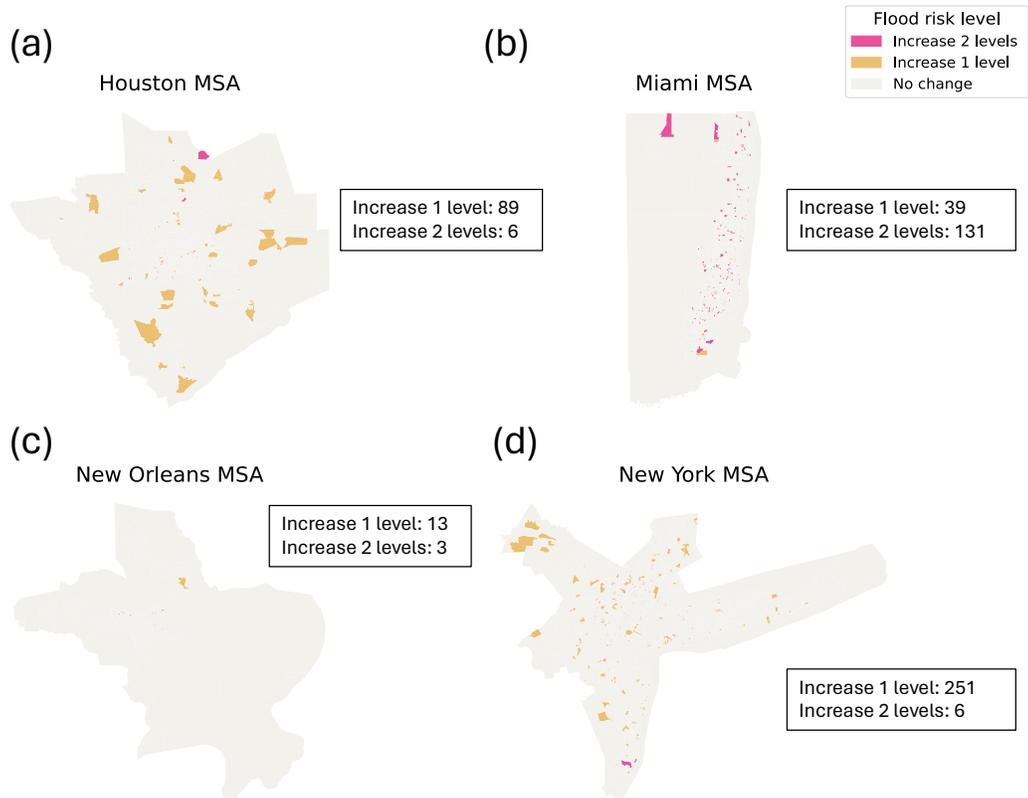

**SF 2. Projected increase in flood risk due to building area expansion across four MSAs**. (a) Houston (b) Miami (c) New Orleans (d) New York. Pink indicates the census block groups with an anticipated increase in flood risk by two levels. Yellow indicates an increase by one level. Gray denotes areas with no expected change. The number for each subplot shows the number of census block groups increase 1 level and 2 levels.